\pdfoutput=1
\documentclass[a4paper,10pt]{article}
\usepackage[bookmarks]{hyperref}
\usepackage{tablefootnote} 
\usepackage[utf8]{inputenc}
\usepackage{float}
\usepackage{listings}
\usepackage{fancyvrb}
\usepackage{graphicx}
\usepackage{urlbreaks}

\hypersetup{
    colorlinks=true,
    linkcolor=black,
    citecolor=black,
    filecolor=black,
    urlcolor=black,
}

\newfloat{Example}{h}{}

\title{Steganography using the Extensible Messaging and Presence Protocol (XMPP)}
\author{Reshad Patuck\footnote{Reshad Patuck: \href{mailto:rap23@kent.ac.uk}{\nolinkurl{rap23@kent.ac.uk}}} , Julio Hernandez-Castro\footnote{Julio Hernandez-Castro: \href{mailto:jch27@kent.ac.uk}{\nolinkurl{jch27@kent.ac.uk}}}\\
\\
\em \small School of Computing, University of Kent,\\
\em \small Cornwallis South Building, Canterbury, CT2~7NF, UK\\}
\date{}

\begin{document}

\maketitle
\hrule

\begin{abstract}
We present here the first work to propose different mechanisms for hiding data in the Extensible Messaging and Presence Protocol (XMPP).
This is a very popular instant messaging protocol used by many messaging platforms such as Google Talk, Cisco, LiveJournal and many others.
Our paper describes how to send a secret message from one XMPP client to another,  without raising the suspicion of any intermediaries.
The methods described primarily focus on using the underlying protocol as a means for steganography, unlike other related works that try to hide data in the content of instant messages.
In doing so, we provide a more robust means of data hiding and additionally offer some preliminary analysis of its general security, in particular against entropic-based steganalysis. 
\end{abstract}

 \hrule

\section{Introduction}

\subsection{Steganography}
Steganography is the science and art of exchanging hidden messages in such a way that nobody, apart from the sender and the intended recipient, suspects the existence of any hidden data.
The word steganography comes from the Greek words \textit{Steganos}  meaning ``covered or protected'' and  \textit{Graphei} meaning ``writing''.

Unlike cryptography, steganography deals exclusively with hiding the information's very presence.
Instead of making the data unreadable, steganography aims to completely hide this data.
An ideal steganographic message will look identical to a regular clean message, and should raise no suspicion \cite{SeeingTheUnseen}.

\subsubsection{History}

The Greek historian Herodotus tells us about the revolt in Ionia (ca. 499 BCE), where Histiaeus shaved the head of his most trusted slave, and tattooed a message onto it.
He then waited for his hair to grow back before sending him to Aristagoras.
Through this communication, Histiaeus sparked a revolt against the Persians \cite[p81]{TheCodebreakers:TheStoryOfSecretWriting}.

Another case of the early use of steganography, also told by Herodotus, was that of Demaratus.
He hid messages by engraving them on wooden tablets, and then covering them with wax \cite[p81-82]{TheCodebreakers:TheStoryOfSecretWriting}.

\subsubsection{Steganalysis}
Steganalysis is ``the art of detecting and decoding messages that have been hidden steganographically'' \cite[p546]{DigitalWatermarkingAndSteganography}.
The most basic goal of steganalysis is to determine whether or not the message may have an embedded ``payload'', at which point the main goal of steganography, which is to communicate secretly, is defeated.
Further goals of steganalysis generally involve determining the means used to hide the data, estimation of the amount of hidden data and, ultimately, recovery of this hidden data.

Steganalysis is generally carried out by a party, commonly named a \emph{warden}.
The \emph{warden} is someone who is free to access all messages exchanged.
Wardens can be categorised \cite{SteganalysisUsingImageQualityMetrics} into two main types:

\paragraph{Passive warden} examines all the data being exchanged, but does not intervene unless something has flagged up its suspicion.

\paragraph{Active warden} changes all data dynamically as the message passes through its network.
This is done to try to remove any hidden content, regardless of whether or not he detects anything suspicious.
This makes perfect sense in some scenarios, and a similar approach is known to have taken place during WWII.
Many letters from soldiers to their families were opened and reworded so as to deliver the same message, destroying any hidden message.
Legend has it that, one such letter was answered with the revealing question: ``Is father dead or deceased?'' \cite{OnTheLimitsOfSteganography}.

\subsection{Extensible Messaging and Presence Protocol (XMPP)}
The Extensible Messaging and Presence Protocol is an XML streaming protocol which allows for the exchange of messages and presence.
\footnote{Presence indicates the status of the user, whether he/she is online, away, or busy.}
The decentralised architecture of XMPP allows each organisation to take control of their own domain, whilst still allowing inter-domain communication.
The protocol relies on XML streams as a medium to transfer data.
This allows any interested party to develop custom functionality over the existing protocol.
This extensibility and decentralisation makes the protocol very flexible and attractive in real life settings.
The multitude of clients, servers and code libraries available makes it one of the more popular choices for companies to base their applications on, making it a \textit{de facto} industry standard.

Some of the larger, and better known deployments of XMPP are:
\begin{itemize}
 \item Google -- The Google Talk service is based on XMPP.
 \footnote{Google has begun phasing out XMPP in favour of their own proprietary hangouts protocol.}

 \item Facebook -- Facebook announced that they would support XMPP for their chat services in 2008 \cite{UsingFacebookChatViaJabber-FacebookDevelopersBlog, FacebookChatAPIFacebookDevelopers}.
 \footnote{In 2009 chat.facebook.com was detected as running a modified version of the ejabberd XMPP server to interface between XMPP clients and Facebook servers.}

 \item Cisco -- Cisco uses XMPP in all of their messaging and presence technologies \cite{JabberTechnology-CiscoSystems}.

 \item LiveJournal -- LiveJournal Talk uses XMPP as a protocol for chat and posting updates to the LiveJournal service \cite{FAQ270-WhatIsLJTalk-LiveJournal}.

 \item  BBC Radio LiveText -- LiveText \cite{BBC-RadioLabs-LiveText} uses an XMPP extension XEP-0124: Bidirectional-streams Over Synchronous HTTP\cite{XEP-0124} to push live text with its radio stream.\footnote{The BBC LiveText service is an example of XMPP used to broadcast data instead of chat.}
\end{itemize}

The XMPP is defined by two `Internet Standard' documents.
These are RFC 6120: XMPP: Core \cite{rfc6120}, which defines the core XML streaming technology, and RFC 6121: XMPP: Instant Messaging and Presence \cite{rfc6121}, which defines the messaging and presence protocols.

\section{Proposed Covert Channels}

Covert channels are so called because they are hidden communication pathways within legitimate communications.
They typically make use of certain properties of a medium (quite frequently high redundancy, but also others) in such a way that it makes the exchange of hidden communication possible \cite{CovertComputerAndNetworkCommunications}.
Covert channels sometimes take advantage of the fact that communication channels have frequently large overheads, and in many implementations need to continuously transfer metadata to keep them open.

XMPP is an example of such a communication protocol, which needs to transfer metadata in each message to keep the channel open.
This metadata is a convenient place to hide information.

\begin{Example}
\begin{Verbatim}[frame=single]
<message from='adam@test.com' to='bart@test.com' type='chat'
id='7df1ddbe'><body>Message.</body></message>
\end{Verbatim}
\caption{An XMPP Message.}
\label{CleanXMPP}
\end{Example}

Example \ref{CleanXMPP} depicts a very simple XMPP message with no hidden data.
We can compare this to the examples shown below, which illustrates the exchange of hidden data.

The covert channels we found and propose are described and analysed below:

\subsection{Type Attribute}

One such place to hide data is in the `\emph{type}' attribute.
This attribute is not used by the server and is passed as-is to the receiver.
This makes it an ideal place to store information.

\subsubsection{Changing the case of the value of the type attribute}
Since the server does not change the value of this attribute, we can change the case of the value in the \emph{type} attribute (from `chat' to `CHAT' or `ChAT') to indicate set or clear bits.
\begin{Example}
\begin{Verbatim}[frame=single]
<message from='adam@test.com' to='bart@test.com' type='CHAT'
id='7df1ddbe'><body>Message.</body></message>
\end{Verbatim}
\caption{One hidden bit coded by the case of the \emph{type} attribute}
\end{Example}

\subsubsection{Changing the value of the type attribute}
Another method of hiding data in this attribute is by using one of the five valid values defined in the XMPP Specification.
This method allows us to encode up to a maximum of 2 bits per message.
However, we recommend using only two of these values `normal' and `chat' to encode just 1 bit per message.
This is because the other two options, `error' and `headline' are not commonly used in chat messages and could easily raise suspicion.
\begin{Example}
\begin{Verbatim}[frame=single]
<message from='adam@test.com' to='bart@test.com' type='normal'
id='7df1ddbe'><body>Message.</body></message>
\end{Verbatim}
\caption{One hidden bit coded in the value of the \emph{type} attribute.}
\end{Example}

\subsubsection{Presence of a type attribute}
Since the XMPP specification (RFC 6121 \cite[subsection 5.2.2]{rfc6121}) does not make the use of this attribute mandatory, we can code another bit in the presence or absence of this attribute in every XMPP message.
\begin{Example}
\begin{Verbatim}[frame=single]
<message from='adam@test.com' to='bart@test.com'
id='7df1ddbe'><body>Message.</body></message>
\end{Verbatim}
\caption{One hidden bit coded in the presence of the \emph{type} attribute.}
\end{Example}

\subsection{id Attribute}

Each XMPP message is allowed to have an \emph{id} attribute.
This can be any string identifying the individual message tag.
It is usually implemented using some sort of counter, numeric or alphanumeric (usually coded in hexadecimal) which is incremented after each message, providing a unique identifier for each message.
This channel too can be steganographically exploited in multiple ways as described below:

\subsubsection{Least significant bits}
The only requirement for these identifiers is that each of them should be unique.
The XMPP specification does not describe an order in which they must occur.
We can use the least significant bits of this identifier as an additional channel to hide data.

An easy way to implement this would be to discard identifier values until the LSBs match the data to be sent.
Simply doing this may look suspicious as the \emph{id} value will change randomly with each message.
A more subtle and much more secure implementation would use the discarded identifier values to send messages to a third party (a decoy XMPP client, who is not participating in the steganographic exchange).
This way, each identifier is used exactly once, making it much harder to detect this type of covert communication.
\begin{Example}
\begin{Verbatim}[frame=single]
<message from='adam@test.com' to='bart@test.com' type='chat'
id='7df1ddbf'><body>Message.</body></message>
\end{Verbatim}
\caption{One hidden bit using the LSB of the \emph{id} attribute.}
\end{Example}

\subsubsection{Case}
Another possible channel that makes use of this \emph{id} attribute employs some of the inherent redundancy in text strings, in this particular proposal, its case.
Since the ID is implemented as a string of characters, we can codify some data as the case of the individual characters in the string.
We can use this to code multiple bits per message.

Using multiple cases in the same \emph{id} is not recommended, as it may flag unnecessary suspicion.
Hence, it is best to stick to one case for the whole \emph{id} and encode just one bit of data in each message.
\begin{Example}
\begin{Verbatim}[frame=single]
<message from='adam@test.com' to='bart@test.com' type='chat'
id='7DF1DDBE'><body>Message.</body></message>
\end{Verbatim}
\caption{One bit coded in the case of the \emph{id} attribute.}
\end{Example}

\subsection{xml:lang Attribute}

Since XMPP relies on XML as its core technology, some of the metadata properties in different XML specifications made their way into XMPP.
The `xml:lang' \cite{LanguageTagsInHTMLAndXML} attribute is one of these properties, used to determine the language the message is written in, such as English (en) or French (fr).
Some channels that can use this property are:

\subsubsection{The presence of the xml:lang attribute}
This attribute is yet another optional attribute which forms part of a valid XMPP message, and one bit of data can be hidden using the presence or absence of this field.
The prototype tool developed while doing this research uses this channel to encode one bit of data.
\begin{Example}
\begin{Verbatim}[frame=single]
<message from='adam@test.com' to='bart@test.com' type='chat'
id='7df1ddbe'><body xml:lang='en'>Message.</body></message>
\end{Verbatim}
\caption{One bit encoded in the presence of the xml:lang attribute.}
\end{Example}

\subsubsection{The value of the xml:lang attribute}
This channel can be extended by assigning a value to the language code in this attribute.
This way, using the redundancy in language codes that represent roughly the same language (en-GB, en-US, etc.), we can encode some data.
\begin{Example}
\begin{Verbatim}[frame=single]
<message from='adam@test.com' to='bart@test.com' type='chat'
id='7df1ddbe'><body xml:lang='en-GB'>Message.</body></message>
\end{Verbatim}
\caption{One hidden bit in the value of the xml:lang attribute.}
\end{Example}

\subsection{Body Contents}
The body of an XMPP message is where the actual content of the message is contained.

Below we list and describe a few of the most practical covert channels employing message contents as a carrier medium.

\subsubsection{Leading and Trailing Space}
The leading and trailing space redundancy is a way of encoding 2 bits of data per message per body element.
This is done by first trimming\footnote{trimming is a string manipulation technique which removes any leading and trailing white-space.} the body of the message, then adding a space character to the beginning and/or end of the message, each encoding one bit of data, in a way very similar to that used by the well-know steganographic tool SNOW \cite{HowSNOWWorks}.
\begin{Example}
\begin{Verbatim}[frame=single]
<message from='adam@test.com' to='bart@test.com' type='chat'
id='7df1ddbe'><body> Message.</body></message>
\end{Verbatim}
\caption{One bit hidden in the leading space.}
\end{Example}

\subsubsection{Replacing Words With Synonyms}
Another possibility for hiding data in the body of a message is replacing words with their synonyms or abbreviations.
This takes advantage of redundancy in the language being used to encode data.

This feature can be implemented as a dictionary of words and their corresponding synonyms.
We can then encode some bits depending on which equivalent word from a list of synonyms we use at a given time.
This implementation can only be detected by performing a heuristic check on the language used in each message \cite{TextSteganographyInChat}, but any such technique will very likely suffer from many false positives and poor overall accuracy.
\begin{Example}
\begin{Verbatim}[frame=single]
<message from='adam@test.com' to='bart@test.com' type='chat'
id='7df1ddbe'><body>Msg.</body></message>
\end{Verbatim}
\caption{One bit hidden by replacing a message with its abbreviation.}
\end{Example}
\subsubsection{Spelling Mistakes}
Spelling mistakes are a common occurrence in quickly typed chat messages.
We can take advantage of some commonly known mistakes to encode some data in whether the word is spelt correctly or not.

This channel's implementation would be quite similar to the synonyms channel described above, in that it would also use a dictionary to encode data and does not have any trivial and highly accurate steganalytic technique.
\begin{Example}
\begin{Verbatim}[frame=single]
<message from='adam@test.com' to='bart@test.com' type='chat'
id='7df1ddbe'><body>Mesage.</body></message>
\end{Verbatim}
\caption{One bit coded by a spelling mistake.}
\end{Example}

\section{Implementing Covert Channels}
\begin{figure}[ht]
 \centering
 \includegraphics[width=60mm]{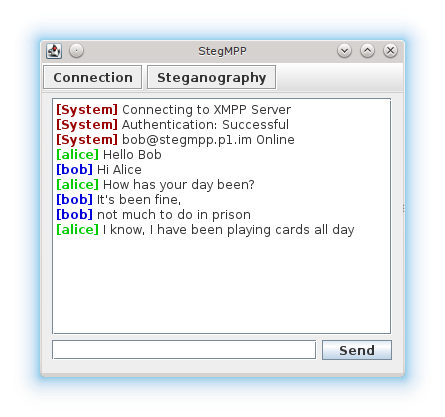}
 \caption{The StegMPP UI}
 \label{StegMPPUI}
\end{figure}

\begin{figure}[ht]
 \centering
 \includegraphics[width=70mm]{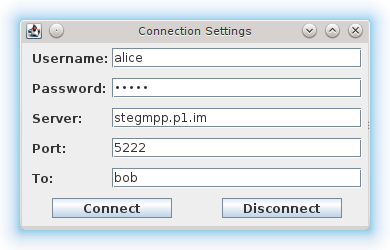}
 \caption{The Connection Settings UI}
 \label{ConnectionUI}
\end{figure}

\begin{figure}[ht]
 \centering
 \includegraphics[width=60mm]{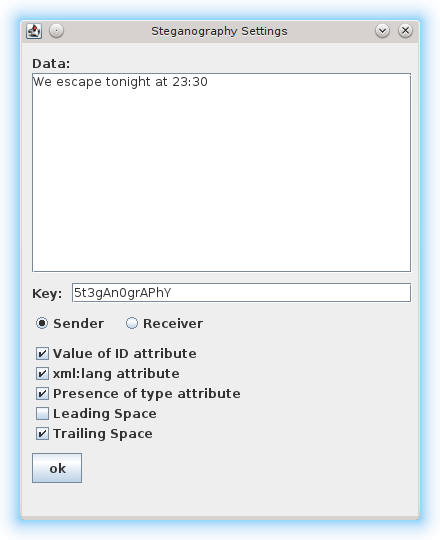}
 \caption{The Steganography Settings UI}
 \label{SteganographyUI}
\end{figure}

We have created a proof of concept tool, called StegMPP\footnote{Available at \url{http://patuck.github.io/StegMPP/}} to implement steganography over some of the covert channels described in previous sections, and to test their capacity and security.
It has a minimalist graphical interface and only implements the bits of XMPP strictly needed to send and receive messages.
Figure \ref{StegMPPUI}, \ref{ConnectionUI} and \ref{SteganographyUI} showcase the user interface of StegMPP.

The steganography sub-system of our tool encrypts and embeds user data into different covert channels.
These are listed below:
\begin{itemize}
 \item Value of the \emph{id} attribute.
 \item Presence of the \emph{xml:lang} attribute.
 \item Presence of the \emph{type} attribute.
 \item Presence of a leading space in the body.
 \item Presence of a trailing space in the body.
 \item \emph{type} attribute case.\footnote{This attribute was finally removed from StegMPP because of known security flaws described in the next section.}
\end{itemize}

\section{Analysis of Covert Channels}

Steganalysis of these covert channels is performed by simulating a warden.
An active warden is not suitable for some of the covert channels described here, because we have found that arbitrarily and/or randomly changing values in some of these channels, would hamper the communication of regular users.
This is why a passive warden is, in the following, preferred to model a realistic setting to monitor the exchanged messages.

Wardens can be further divided into two kinds, depending on whether they take into account previous states of the protocol:

\subsection{Stateless Warden}
A stateless warden is one that treats every message independently and does not take into account any of the previously seen messages.
This type of warden is easy to implement, as it only needs to compare each message to a static list of definitions.

However this type of warden is only able to detect hidden data that is not a valid XMPP message.
Channels like the case of the `type' attribute can be detected as the RFC \cite[subsection 5.2.2]{rfc6121} specifies certain values, all of which are in lower case.
This makes finding hidden communications trivial as any message with upper case values in its `type' attribute can be immediately categorised as suspicious.

\subsection{Stateful Warden}
A stateful warden is more powerful than a stateless warden in many ways.
It keeps track of the previous messages and compares all new messages against a list of previous messages.
This makes catching hidden content easier as the warden is not matching against a list of definitions but against previous states of the protocol.
\begin{table}
 \centering
 \begin{tabular}{ p{6cm}  p{2cm}  p{2cm} }
  \hline
  Covert Channel & Stateless Warden & Stateful Warden  \\ \hline
  Presence of type Attribute        & no  & yes  \\
  Value of ID Attribute            & no  & yes\tablefootnote{Can be undetectable if discarded id's are used to send other clients messages.}   \\
  Presence of xml:lang Attribute    & no  & yes  \\
  Leading Space                     & no  & no   \\
  Trailing Space                    & no  & no   \\
  Case of type attribute\tablefootnote{This attribute was removed from StegMPP as it was detectable even by a stateless warden.} & yes & yes  \\ \hline
  \end{tabular}
  \caption{Covert channels and their detectability.}
  \label{TestedCovertChannels}
\end{table}

\section{Results}
The results of simulating both types of warden show that most of the channels implemented by StegMPP are immune to detection by a stateless warden.
It also shows that some of these channels can be detected by a carefully crafted stateful warden.
However implementing a stateful warden to monitor all the connections on a busy XMPP server is not trivial and requires significant computational resources and being able to deal with a potentially very high number of false positives.

Table \ref{TestedCovertChannels} on page \pageref{TestedCovertChannels} lists the covert channels (each hiding one bit per message) tested, and their security against both stateless and stateful wardens.

\subsection{Statistical Analysis}

We used \emph{ent}, a tool implementing a battery of statistical tests, to compute the entropy per byte and find whether there were statistically significant differences between a normal (clean) XMPP session and one with covert channels being employed to exchange some data.
The results of this analysis can be found in Table \ref{EntropyOfChannels} on page \pageref{EntropyOfChannels}.

\begin{table}
 \centering
 \begin{tabular}{p{6cm} p{2cm} p{2cm}}
  \hline
  Channel				& Entropy	& Difference \\
  \hline
  Control				& 4.976370	& \\
  Presence of type Attribute		& 4.974257	& -0.002113 \\
  Value of ID Attribute			& 4.980152	& ~0.003782 \\
  Presence of xml:lang Attribute	& 4.989826	& ~0.013456 \\
  Leading Space				& 4.975970	& -0.000400 \\
  Trailing Space			& 4.975430	& -0.000940 \\
  \hline
 \end{tabular}
 \caption{Comparison of entropy of covert channels.}
 \label{EntropyOfChannels}
\end{table}

A detailed inspection of Table \ref{EntropyOfChannels} clearly reveals that any statistical steganalysis of the proposed techniques based on entropic measures would have to be extremely precise due to the minuscule differences between the values corresponding to the clean (Control) exchange and the rest of the hidden ones.

An additional indication of this difficulty can be observed in the fact that not always the entropy corresponding to the channels with hidden contents is higher that that of the clean one.  This is frequently the case in different steganographic settings, but in here we have alternating signs in the last column (Difference).

That said, these figures also show that the covert channel based on the presence of xml:lang attribute seems to be orders of magnitude easier to detect than the rest, so its usage cannot be recommended in high security environments.

This, by itself, is of course no proof of security, but at least shows that the techniques proposed in this article and implemented in the accompanying tool modify the cover media only very slightly, giving us some indication that any future entropy based steganalysis will need large amounts of data and computational power to succeed.

\section{Conclusion}

A new method for steganographic communications using XMPP is proposed for the first time in this work.
A number of new and promising covert channels are discussed in detail.
Most of these channels take advantage of the redundancy in the underlying protocol, instead of the contents of the message.
Simulations show that most of these channels can go undetected by a stateless warden, but may be detected by a stateful warden.
We in addition propose some subtle ways to harden a subset of these channels against attacks by stateful wardens.

Our proof of concept tool, StegMPP, comes with many of the proposed functionalities described in this paper, and is freely available for download at: \url{http://patuck.github.io/StegMPP/}

\subsection{Future Works}

In order to develop a more secure implementation, we propose further research into how to harden steganalysis for a stateful warden.
One of the ways we may be able to accomplish this is by alternating the channels we use in each message.
We believe this idea is promising and plan to study it in more detail in the near future.
We would also like to find a way to improve the efficiency of a stateful warden capable of detecting these covert channels, and develop a steganalytic tool to detect the presence of hidden messages in XMPP messages.
This tool could potentially be of use to institutions, large corporations and governmental organisations, particularly for early detection of insider threats and data leakages.




\nocite{AProposalOnInformationHidingMethodsUsingXML}
\nocite{SteganographyAndDigitalWatermarking}
\bibliographystyle{unsrt++}
\bibliography{Article}

\end{document}